%% file: firstpaper.tex
\documentclass[usenatbib,usegraphicx]{mn2e}

\usepackage{ifthen,natbib,graphicx} \usepackage{comment} 

\usepackage{amsmath,times}
\usepackage{url,astrojournals}

\newboolean{includeversions} \setboolean{includeversions}{false}

\newboolean{blackandwhite}

\include{natlatex_macros}
\renewcommand{\mcrx}{{\sc sunrise}} 
\renewcommand{\hii}{{H{\sc II}}} 
\newcommand{\hi}{{H{\sc I}}} 

\begin{document}

\title [Dust Attenuation in Spiral Galaxies]{Dust Attenuation in Hydrodynamic Simulations of Spiral Galaxies}

\author[Rocha et al.]  {Miguel
Rocha,$^1$\thanks{E-mail: mrocha@scipp.ucsc.edu} 
Patrik Jonsson,$^1$\thanks{E-mail: patrik@ucolick.org} Joel
R. Primack,$^{1,2}$\thanks{E-mail: joel@scipp.ucsc.edu} and T.J. Cox$^3$\thanks{E-mail: tcox@cfa.harvard.edu} \\ $^1$Santa Cruz
Institute for Particle Physics, University of California, Santa Cruz, CA
95064\\ $^2$Department of Physics,
University of California, Santa Cruz, CA 95064  \\
$^3$Harvard-Smithsonian Center for Astrophysics,Cambridge, MA 02138}

\maketitle

\begin{abstract} 
We study the effects of dust in hydrodynamic simulations of spiral
galaxies when different radial metallicity gradients are
assumed. \mcrx, a Monte-Carlo radiative-transfer code, is used to make
detailed calculations of the internal extinction of disk galaxies
caused by their dust content.

\mcrx \ is used on eight different Smooth Particle Hydrodynamics
(SPH) simulations of isolated spiral galaxies. These galaxies vary
mainly in mass and hence luminosity, spanning a range in luminosities
from $-16$ to $-22$ magnitudes in the $\mathrm{B}$ band. 
We focus on the attenuation in different wavelength bands as a
function of the disk inclination and the luminosity of the models, and
compare this to observations.

Observations suggest different metallicity gradients for galaxies of
different luminosities. These metallicity gradients were explored in
our different models, finding that the resulting dust attenuation
matches observations for edge-on galaxies, but do not show a linear
behaviour in $\log \mbox{axis ratio}$ as some observations have
suggested. A quadratic law describing the dependence of attenuation on
inclination, as proposed by more recent observations, reconciles the
attenuation of the simulations at intermediate inclinations with
observations. We also compare the total infrared-to-ultraviolet flux
ratios for the simulated galaxies with those of the SINGS sample and
find general agreement. Finally we compare our results with those
from simpler models that do not take into account structure such as
spiral arms, finding that the inclusion of sub-structure on the
size scale of spiral arms does not change conclusions about the
attenuation dependence on inclination or wavelength.
\end{abstract}


\section{Introduction}

Studies of dust extinction in spiral galaxies play an important role
in this time of precision cosmology. It is remarkable how, despite the
fact that distances have always been one of the most challenging
measurements in astronomy, we are now able to obtain accurate
cosmological distances to galaxies and galaxy clusters. This is
possible because of relations like the Tully-Fisher relation
(hereafter TF) and the Fundamental Plane, which relate observable
parameters of galaxies, such as $\mathrm{H_I}$ velocity profile widths
and central velocity dispersions, to their luminosity and
radius. Given that these methods are independent of redshift, they are
also important in determining the Hubble constant
\citep{Freedmanetal01} and peculiar velocities of galaxies (i.e., the
difference between Hubble velocities and redshift velocities). With
this, one can also obtain a potential \citep{Bertschingeretal90} that
can be used to compare the observed galaxy distributions and
kinematics to cosmological predictions \citep[e.g.][]{Dekeletal99}.
Nevertheless, these valuable relations are subject to extinction
corrections that are important to take into account. The scatter in
the TF relation is increased if the extinction correction that depends
on the disk inclination is not well understood. Even more so, the
slope of the TF relation may be affected by corrections for the
luminosity dependence of extinction. The dependencies of the TF
relation on such corrections are demonstrated in detail in
\citet[][hereafter G95]{Giovanellietal95}.

Furthermore, hierarchical models of galaxy formation predict that the
disks of spiral galaxies form from the inside out, having younger
stellar populations and lower metallicity in the outer disk. Thus,
quantifying the gradients in the metallicity distribution of galaxies
with different luminosities is important in order to compare the
predictions of hierarchical models with observations. This can be done
by looking at the effects that different metallicity gradients have on
the attenuation of disk galaxies, and comparing the results with
observations.

Another motivation for studying the attenuation of dust in the disk of
spiral galaxies is to gain insight into how the dust is
distributed. The attenuation vs. inclination relations of a galaxy
completely free of dust and a completely opaque galaxy in which the
obscuring dust is confined to a thin layer in the mid-plane of the
disk are difficult to distinguish, except when seen close to
edge-on. This is one of the reasons why determining the degree of dust
attenuation in a galaxy is difficult, and has led to a long-standing
debate about the extent to which galaxy disks are transparent or
opaque \citep{Disneyetal89,Bursteinetal91,Bosmaetal92,Disneyetal92,
Valentijnetal94,Kylafisetal01,Holwerdaetal05}. Further studies
\citep{Moriondoetal98,Xilourisetal99} indicate that central regions of
the disk may be opaque while outer regions are
transparent. Observations reveal that the surface intensity in spirals
falls off exponentially with radius \citep{dV48} while in the vertical
direction both the stars \citep{Kruitsearle81} and the gas and dust
\citep{Cowiesongalia86,Kylafisbahcall87} fall off roughly
exponentially with height. Exponential models of galaxies with equal
dust and stellar scalelengths have previously been used to study dust
attenuation \citep{Disneyetal89}. \citet[][hereafter
X99]{Xilourisetal99} fit exponential models to observations of seven
spiral galaxies seen very close to edge-on (NGC 4013, IC 2531, UGC
1082, NGC 5529, NGC 5907, UGC 2048, and NGC 891), and revealed that in
general these large spirals possess a dust scalelength that is about
1.4 times larger than that of the stars, and a vertical dust
scaleheight that is about half that of the stars. These data
indirectly suggest a metallicity gradient that will have different
effects on the shape and amplitude of the attenuation vs. inclination
relation.  Consequently, exploring this relation can lead to
conclusions about the radial distribution properties of the dust in
the disk of spiral galaxies.

Other observations have quantified the metallicity gradient of spiral
galaxies by different methods.  \citet[][hereafter
Z94]{Zaritskyetal94} provides the metallicity gradient of 39 disk
galaxies, by examining the properties of \hii\ regions within the disk
of these galaxies. They found an average metallicity gradient of $-0.05$
dex/kpc, although this value varied widely from $0.00$ to $-0.20$
dex/kpc. They found a correlation between abundance gradients and Hubble
type only when the gradients are expressed in terms of dex/kpc as
opposed to dex/isophotal radius.

Based on the fact that colour gradients in disk galaxies are
due in most part to the gradients in age and metallicity
\citep{belldejong00}, \citet[][hereafter MA04]{MacArthuretal04}
provided results of metallicity gradients from observations of optical
and NIR galaxy colour gradients in a sample of 172 galaxies with
different luminosities. They found that the metallicity gradients in
their sample were steeper than those of Z94 for galaxies fainter than
$\sim -20$, and shallower for galaxies brighter than $\sim -20$, in
the K-band.

G95 sampled more than 1700 galaxies for which I-band CCD images,
redshifts, and \hi\ line widths were available. Their conclusions were
that their attenuation as a function of inclination could be well
fitted by a linear function of $\log (a/b)$, where $a/b$ is the axis
ratio. From sampling 87 spiral galaxies within the Ursa Major and
Pisces Clusters, \citet[][hereafter T98]{Tullyetal98} reached
conclusions similar to those of G95, with their data fitting well a
linear relation. \citet[][hereafter M03]{Mastersetal03} based their
results on data taken from 3035 galaxies from the 2 Micron All-Sky
Survey \citep[2MASS][]{jarrettetal01, jarrettetal03} for which I-band
photometry and redshifts were available, and concluded that a simple
linear law was not adequate for describing the attenuation dependence
on inclination observed in their sample. Thus, M03 adopted a bilinear
law fitting separately high and low inclinations systems as separated
by $\log (a/b) = 0.5$. Looking for even less linear shapes that fit
their observed colour dependence on inclination, M03 also tried a
quadratic law. The results provided by M03 and G95 cover only the
infrared and near infrared range of the spectrum.

Here we use the capabilities of \mcrx, a Monte Carlo
radiative-transfer code \citep{jonsson06sunrise}, to perform
``simulated observations'' of dust attenuation in the disks of eight
different hydrodynamic simulations of spiral galaxies, having
different luminosities and sizes, while assuming the different
metallicity gradients provided by MA04, X99, and Z94. Performing these
``simulated observations'' at different inclinations for each of the
galaxy models, we obtain the attenuation dependencies on inclination
and luminosity for each of the assumed metallicity
distributions. These results are then compared to the observations
made by M03, G95, and T98. By finding which simulations best fit the
data, it is determined which metallicity gradients are the most
appropriate for these galaxy models.

The main difference between this work and previous ways of quantifying
the attenuation produced by dust in galaxies through radiative
transfer calculations is that our calculations are performed over an
arbitrary geometry specified by the hydrodynamic simulations. Thus our
galactic models resolve structure such as spiral arms that in previous
radiative transfer models, either by analytical approximations
\citep{Byunetal94,Silvaetal98,Xilourisetal99,BaesDejonghe01,Tuffsetal04}
or by Monte Carlo calculations \citep{deJong96, Kuchinskietal98,
Ferraraetal99, bfg96, Bianchietal00, MatthewsWood01, Pierinietal04},
is not considered. Nevertheless, the structure resolved in our
simulations is not small enough to include individual star forming
regions, which affects the resulting attenuation. We compare our
results with previous models to examine if the inclusion of structures
such as spiral arms makes a difference in the dependence of
attenuation on inclination and wavelength. Finally we have extended
the triple exponential model of \citet{Disneyetal89} to explore how
the dependence of attenuation on inclination changes with different
dust-to-star scalelengths ratios.

The structure of this paper is as follows: Section 2 is a presentation
of the hydrodynamic galaxy models. In section 3, a description is
given of how the simulated observations were made with \mcrx. In
section 4 the results are compared to real observations. In section 5
we compare to previous models of galactic disks without structure such
as spiral arms, and explore how the dependence of attenuation on
inclination changes with different dust-to-star scalelength
ratios. Section 6 provides a summary and lists the conclusions.

\section{The Hydrodynamic Galaxy Models}

Radiative-transfer calculations were made for eight different Smooth
Particle Hydrodynamics (SPH) simulations of isolated galaxies. These
simulated galaxies were created using a version of the code {\sc
gadget1} which enables entropy conservation
\citep{Springeletal01,SpringelHernquist02}, as part of a comprehensive
work aimed at studying galaxy mergers \citep[][]{tjthesis,
Coxetal06,Coxetal07,pjetal05attn,jonsson06sunrise}.

The set of isolated galaxies for which calculations of dust extinction
were made includes four models that approximate late-type spiral
galaxies and four less gas-rich spiral galaxies (the G-series). The
late-type spirals group comprises an Sbc type galaxy, a more massive
and bigger Sbc (Sbc+), a less massive and smaller Sbc (Sbc-), and an
Sc type. The G-series have been created spanning a wide range of
masses and having average properties for their mass. Table
\ref{galaxymodels} lists the characteristic of each model. The details
of these galaxy models are described in \citet{Coxetal06} and
\citet{pjetal05attn}. Here we mention some of their important
features. 

The galaxies contain a stellar and gaseous disk, which are
rotationally supported, a spheroidal stellar bulge, and a massive dark
matter halo. A model for supernova feedback, as described in
\citet{Coxetal06}, is included and is crucial for the stability
of the gas disk and for preventing excessive star formation. The
adopted star formation prescription is based on a Schmidt law
\citep{Schmidt59} and is consistent with the Kennicutt law
\citep{Kennicutt98}. Schemes for radiative cooling and metal
enrichment have also been implemented in these models. Our models do
lack age gradients, something that would be necessary for comparing
our results with the predictions of hierarchical models of galaxy
formation.

\begin{table*}
\begin{minipage}{160mm}
\caption{The galaxy models used for the calculations of dust attenuation.}
\label{galaxymodels} 
\begin{tabular}{lccccccccccccc}
\hline Model & $M_{\mbox{vir}}$$^\mathrm{a}$ & $M_{b}$$^\mathrm{b}$ &
$R_{d}$$^\mathrm{c}$ & $Z_{d}/ R_{d}$$^\mathrm{d}$ &
$R_{g}/R_{d}$$^\mathrm{e}$ & $f_{g}$$^\mathrm{f}$ &
$f_{b}$$^\mathrm{g}$ & $R_{b}$$^\mathrm{h}$ &
$V_{\mbox{rot}}$$^\mathrm{i}$ & $Z_{1.3}$$^\mathrm{j}$ &
$N_g$$^\mathrm{k}$ & Age $^\mathrm{l}$ & $\tau$ $^\mathrm{m}$ \\ &
($\Msun$) & ($\Msun$) &(kpc) & & & & & ($\kpc$) & $\kps$ & (Z$_\odot$)
& $(\mathrm{x}10^4)$ & (Gyr) & (Gyr) \\ \hline Sbc+ & $9.28\cdot10^{11}$ &
$1.56\cdot10^{11}$ & 7.0 & 0.125 & 3.0 & 0.52 & 0.10 & 0.60 & $210$ &
$1.12$ & 3 & 13.9 & 110 \\ Sc & $8.90\cdot10^{11}$ &
$1.12\cdot10^{11}$ & 4.7 & 0.2 & 4.0 & 0.69 & 0.00 & 0.00 & $196$ &
$1.00$ & 3 & 13.8 & -13\\ Sbc & $8.12\cdot10^{11}$ &
$1.03\cdot10^{11}$ & 5.5 & 0.125 & 3.0 & 0.52 & 0.10 & 0.45 & $195$ &
$1.00$ & 3 &13.9 & -106\\ G3 & $1.16\cdot10^{12}$ & $6.22\cdot10^{10}$
& 2.8 & 0.125 & 3.0 & 0.20 & 0.14 & 0.37 & $192$ & $1.00$ & 5 & 14.0 &
10 \\ Sbc- & $3.60\cdot10^{11}$ & $4.98\cdot10^{10}$ & 4.0 & 0.125 &
3.0 & 0.52 & 0.10 & 0.40 & $155$ & $0.70$ & 3 & 13.7 & 124\\ G2 &
$5.10\cdot10^{11}$ & $1.98\cdot10^{10}$ & 1.9 & 0.2 & 3.0 & 0.23 &
0.08 & 0.26 & $139$ & $0.56$ & 3 & 14.0 & 8.2 \\ G1 &
$2.00\cdot10^{11}$ & $7.00\cdot10^{9}$ & 1.5 & 0.2 & 3.0 & 0.29 & 0.04
& 0.20 & $103$ & $0.40$ & 2 & 11.5 & 3.7 \\ G0 & $5.10\cdot10^{10}$ &
$1.60\cdot10^{9}$ & 1.1 & 0.2 & 3.0 & 0.38 & 0.01 & 0.15 & $ 67$ &
$0.28$ & 1 & 8.7 & 1.4 \\ \hline
\end{tabular}

\medskip
 $^\mathrm{a}$Virial mass.\\ $^\mathrm{b}$Baryonic mass. \\
 $^\mathrm{c}$Stellar disk scalelength. \\ $^\mathrm{d}$Ratio of
 stellar-disk scaleheight and scalelength. \\ $^\mathrm{e}$Ratio of
 scalelengths of gas and stellar disks.\\ $^\mathrm{f}$Gas fraction
 (of baryonic mass).\\ $^\mathrm{g}$Bulge fraction (of baryonic
 mass).\\ $^\mathrm{h}$Bulge scale radius.\\ $^\mathrm{i}$Circular
 velocity.\\ $^\mathrm{j}$Metallicity at 1.3 scalelengths from the
 centre (gas and stars), from Z94. \\ $^\mathrm{k}$Number of gas
 particles.\\ $^\mathrm{l}$Age of oldest stars (formation time of
 bulge and oldest disk stars). \\ $^\mathrm{m}$Exponential time
 constant of the star formation rate for the disk stars.

\end {minipage}
\end{table*}

\section{The Radiative Transfer Calculations}

\subsection{\sc sunrise}

\mcrx \ is a high-performance code designed to combine a full
radiative-transfer model with high-resolution hydrodynamic simulations
of either isolated or interacting galaxies; see
\citet{jonsson06sunrise} for a detailed description of how \mcrx\
achieves this goal. \mcrx \ first assigns an initial spectral energy
distribution (SED) for the stellar particles in the hydrodynamic
simulations. Assuming that the density of dust traces the density of
metals, it then solves the radiative-transfer problem by the
Monte-Carlo method in the given arbitrary geometry for a number of
wavelengths, and finally, it interpolates to obtain a resulting full
SED after absorption and scattering. The SEDs used are from
Starbust99\footnote{\url{http://www.stsci.edu/science/starburst99/}}
v5.0 \citep{Leithereretal99} with Kroupa IMF. In post-processing,
\mcrx \ calculates total magnitudes and generates images for the
specified optical filter bands, as well as the total absorbed
bolometric luminosity, over all wavelengths. In this way \mcrx \ can
be used to make simulated observations that can be easily compared to
actual observations.  \mcrx \ is free
software\footnote{\url{http://sunrise.familjenjonsson.org}}.

For this investigation, calculations have been done using eight
different filters ($\mathrm{U, B, V, R, I, J, H}$ and $\mathrm{K}$)
and thirteen cameras at different inclinations. Six cameras were
placed above the plane of the disk at inclination angles 0, 35, 51,
63, 74, and 85 degrees.  Another six were placed at the same
inclinations but on the opposite side of the disk and at a 90 degree
azimuthal angle to the first set.  One more camera was placed exactly
edge-on. These calculations were done for two simulation times, 0.5
Gyr and 1 Gyr. This allows enough time for the simulation to relax but
is not so late that a significant amount of gas is consumed. For each
inclination, the results were averaged over the two azimuthal angles
and the two snapshots.  For simplicity, the dust model used in all our
models regardless of size was the $R=3.1$ Milky-Way model by
\citet{weingartnerdraine01}.

\subsection{Star-Formation History and Metallicity Distributions}

Given that the metallicity and the age are required to pick the SED of
a stellar particle, two important values that are given as inputs to
\mcrx \ are the star-formation history and the metallicity
distribution of the stars present when the simulation is started.

In all the calculations made for this paper the bulge stars are
assumed to have formed in an instantaneous burst, while the disk has
had an exponentially declining star-formation rate starting at the
time of bulge formation and leading up to the start of the
simulation. The age of the bulge and stars at the beginning of the
simulation, and the time constant $\tau$ of the exponential decline
used for each of the galaxy models are given in Table
\ref{galaxymodels}. The time constant $\tau$ for each galaxy was
determined by fitting an exponential star formation rate history to
the following constraints: the present-day star-formation rate must
match the star formation rate in the hydrodynamic simulations, the
total stellar mass must match that in the simulated models, the
average stellar age must match the data in MA04 (as well as possible),
and finally no stars can be older than a Hubble time. In the case of
the Sc galaxy, this means a marginally growing star-formation rate (at
the onset of star formation 13.8 Gyr ago, the star formation rate was
37\% of the present-day SFR).  This is not very surprising, since a
large, gas rich spiral today cannot have sustained its large star
formation rate for a long time without exceeding its current stellar
mass, it's just being assembled later than average. Effectively, the
very large numbers (positive as well as negative) just mean a constant
star formation rate history.

The metallicity of the gas and stars at the start of the simulation
is, like the surface densities, assumed to decline exponentially with
radius in the disk. The metallicity of the gas particles is assumed to
be the same as that of the stars and thus the same dependence applies
for the gas metallicity as a function of radius. \citet{Dwek98}
concluded that the formation of dust grains in the interstellar medium
is, in good approximation, linearly proportional to the density of
metals, with a proportionality constant of 0.4. If that is the case,
the dust density $d(r)$ would also decline exponentially from the
centre as is observed by X99, and be described by a relation of the
form

\begin{equation}
\label{dustdist} d(r) = 0.4g(r)z(r) = d_o e^{ - r/R_g } 10^{ - G{\rm{ }}r}  = d_o e^{ -r/R_d}  \ ,
\end{equation}
with $R_g$ and $R_d$ being the gas and dust scalelengths respectively,
$d_o = 0.4 z_o g_o$ being the central density of dust, and G being a
constant that determines the gradient of a metallicity distribution
expressed in a decimal exponential manner, as is usually given
(Z94, MA04).  The observations of X99, as described in Section 1,
suggest that the dust scalelength of big spiral galaxies is about 1.4
times greater than the stellar scalelength. Hence, we can define the
``Xilouris gradient'' $G_\mathrm{X}$ as the gradient that results in a
ratio of dust scalelength to stellar scalelength equal to 1.4. Recalling equation
\ref{dustdist} we have that
\begin{equation}
e^{ - {r}/{{R_g }}} 10^{ - G_\mathrm{X} {\rm{ }}r}  = e^{ - {r}/{{R_d }}}  = e^{ - {r}/{{1.4R_s }}} \ ,
\end{equation} 
or 
\begin{equation}
G_\mathrm{X}  = \left(\frac{1}{1.4R_s} - \frac{1}{R_g}\right)\log e \ .
\end{equation}
The Xilouris gradient $G_\mathrm{X}$ for each of the galaxy models is shown in Table \ref{testedgrads}.

X99 observed only big spiral galaxies, so $G_\mathrm{X}$ is not
necessarily the best metallicity gradient assumption for our set of
smaller galaxies. MA04 (Fig. 26) provides metallicity gradients as a
function of luminosity for a sample of 127 galaxies, from which we
have obtained another set of gradients that were applied to the three
smallest galaxies, G2, G1 and G0. Table \ref{testedgrads} shows these
last gradients and summarises all the gradients that were tested in
the simulations.

\begin{table}
\caption{Tested Metallicity Gradients.}
\label{testedgrads}
\begin{tabular}{cccc} 
\hline
Galaxy & Z94      & X99 $(G_\mathrm{X})$   & MA04    \\
       & $(-dex \ kpc^{-1})$ & $(-dex \ kpc^{-1})$ & $(-dex \ kpc^{-1})$ \\
\hline
Sbc+& 0.05& \textbf{0.023}& - \\Sc & 0.05 & \textbf{0.043}&- \\Sbc&0.05&\textbf{0.030}&- \\G3&0.05&\textbf{0.058}&- \\ Sbc-&0.05&\textbf{0.041}&- \\G2&0.05&0.086&\textbf{0.04} \\G1&0.05&0.112&\textbf{0.05} \\G0&0.05&0.148&\textbf{0.06}\\
\hline
\end{tabular}

The adopted metallicity gradients in our models are shown in bold.
\end{table}

It is important to note that the central metallicity of the galaxies
depends on the gradient in use. This is because the metallicity given
in Table \ref{galaxymodels} is at 1.3 scalelengths from the
centre.

The advantage of producing simulated observations with \mcrx \ is that
the results of the calculations can be directly compared to
observations, which will be done in the next section.

\section{Results and Comparison with Observations}

\begin{figure} 
\begin{center}\includegraphics[width=80mm]{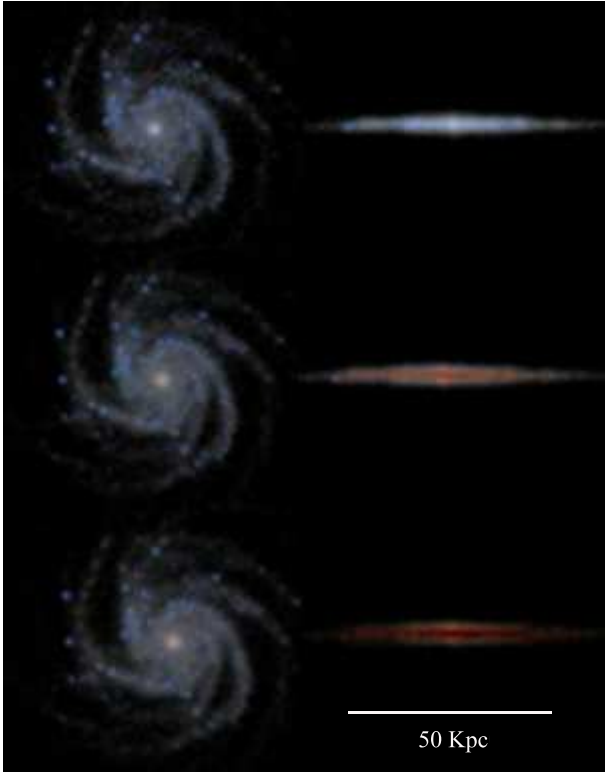}\end{center}
\caption{ Colour composite images (I,V,B bands) of the Sbc model with no dust, assuming the Xilouris gradient, and assuming constant metallicity for top, middle, and bottom respectively. We can see a significant difference between these three cases when edge-on. }
\label{sbcedge}
\end{figure}

\begin{figure} \begin {center} \includegraphics[width=80mm]{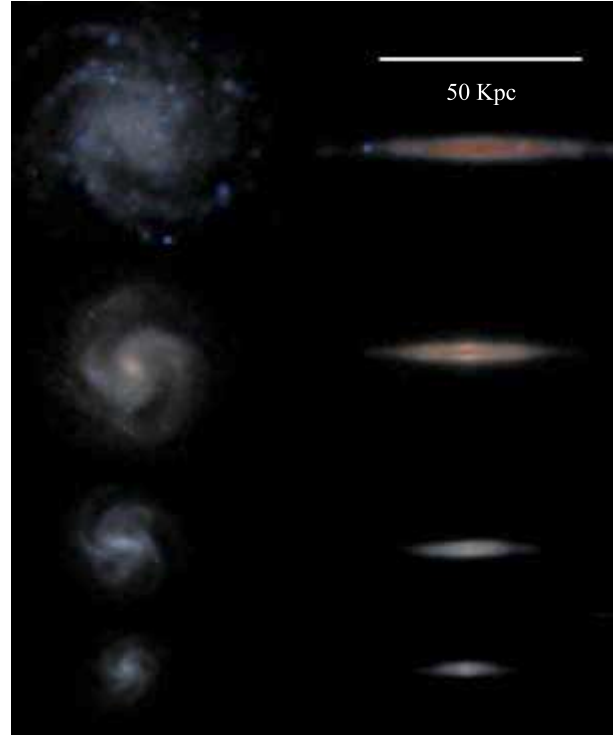} \end {center}  \caption{ Face-on and edge-on colour composite images  (I,V,B bands) of the Sc, G3, G2 and G1 galaxy models from top to bottom respectively. The scale is the same so that the differences in size can be visualised. The G0, which is not shown here, is even smaller and less luminous than the G1 (see Table \ref{galaxymodels}). Notice how the strength of the dust lane and the reddening decreases with the size (luminosity) of the galaxy, and is unnoticeable for the G1 and G2. }
\label{scandgs}
\end{figure}

Before analysing any data, we can anticipate some of the results by
simply looking at the colour composite images obtained in our
calculations. Figure \ref{sbcedge} clearly shows the effects of having
or not having a metallicity gradient when the Sbc model is seen
edge-on, while Figure \ref{scandgs} shows how the metallicity, and
hence the strength of the dust lane and attenuation, decrease with
luminosity. Note that \citet{Dalcantonetal04} found that galaxies with
rapidly rotating disks (i.e. $V_{\mathrm{rot}} > 120\kps$) have well
defined dust lanes, while galaxies with lower rotation velocities do
not show evident dust lanes. Figure \ref{scandgs} shows that it is
hard to see any dust lanes in our G2 and smaller galaxy models, which
have rotation velocities of less than $139\kps$ (See Table
\ref{galaxymodels}). 
   
\subsection{Observations of Disk Attenuation.}

Some of the results of G95, T98, and M03 will be used to compare the
data obtained from our simulations with observations. These authors
have quantified the extinction effects in the disks of galaxies as a
function of inclination, with an aim to better understand the
necessary extinction corrections to the TF relation. The extinction
effects of dust are normally quantified by measuring the attenuation,
which refers to the decrease in flux resulting from photons being
scattered and/or absorbed. This is defined as
\begin{equation}
\label{Att} 
A_\lambda = m_\lambda^0-m_\lambda =
- 2.5 \log\left(\frac{f_\lambda^0}{f_\lambda} \right),
\end{equation}
with $m_\lambda^0$ and $f_\lambda^0$ being the intrinsic magnitude and
flux in the absence of dust respectively. In the simulations, it is
simple to find the attenuation dependence on inclination of a galaxy
by simply computing the attenuation, as in equation \ref{Att}, for a
number of different lines of sights at different inclinations. In real
systems, it is not easy to determine the actual inclination of the
disk and it is impossible to know the intrinsic magnitude or flux by
direct measurement, hence more practical ways of defining inclination
and attenuation are required.

The technique used (G95, T98 and M03) to find the inclination is by
first determining an isophotal radius (a radius of constant surface
brightness) in the disk of the observed galaxies, and then finding the
ratio between the semi-major and the semi-minor axis of the isophotal
ellipse, which is related to the actual inclination by
\begin{equation}
\label{inc_ratio} \cos i = \sqrt {\frac{{({b}/{a})^2  - q_{}^2 }}{{1 - q_{}^2 }}} \ ,
\end{equation}
where $a$ and $b$ are the semi-major and semi-minor axis respectively,
and $q$ is the axial ratio when the system is viewed edge-on. Having
this and given that surface brightness is proportional to the flux, one
could expect that, as a simple approximation, the attenuation relative
to face-on as a function of axial ratio would follow a relation of the
form
\begin{equation}
\label{att_ratio} 
A_\lambda   = \gamma _\lambda  \log\left(
\frac{\mbox{face-on area} = \pi a^2}{\mbox{inclined area} = \pi
  ab}\right) = 
  \gamma _\lambda  \log\left(\frac{a}{b}\right) , 
\end{equation}
where $\gamma_\lambda$ is an attenuation parameter as a function of
wavelength $\lambda$ and luminosity. Of course this would be
completely valid only if the surface brightness of the system were
constant. G95, T98 and M03 fitted their observations with a relation
of the form of equation \ref{att_ratio} and with that obtained the
most appropriate value of $\gamma_\lambda$.

Since it is not possible to know the intrinsic magnitudes and these
observers were only interested in the dependence on inclination of the
attenuation in a given band, the observations used for their fits were
those of the $\lambda-\mathrm{K}$ colour (where $\lambda$ can be any
given band).  This is because dust is optically thin in the near
infrared part of the spectrum, and thus the magnitude in $\mathrm{K}$
should be approximately constant with inclination. Thus, the main
technique used for these observations is to find the $\gamma_\lambda$
that best fits their observations of the dependence of the colour
$\lambda-\mathrm{K}$ on the axial ratio $a/b$. Under the
assumption that extinction effects in the K-band are negligible and
independent of inclination, the dependence of $\lambda-\mathrm{K}$
colour on inclination would then represent the actual attenuation
dependence on inclination for the given band $\lambda$.

T98 assumed that the extinction effects in the K-band, although small,
were not entirely negligible. They concluded that their results of
attenuation as a function of inclination were well fit by a linear
relation in logarithmic space that followed the form of equation
\ref{att_ratio} (see Fig 2 in T98), with a slope $\gamma_\lambda$ that
is dependent on luminosity. Their adopted equations for $\gamma_\mathrm{B}$ and
$\gamma_\mathrm{I}$ were
\begin{equation}
\label{gammaB}\gamma _\mathrm{B}  =  - 0.35(15.6 + M_\mathrm{B}  + 5\log h_{80} )
\end{equation}
\begin{equation}
\label{gammaI} \gamma _\mathrm{I} =  - 0.20(16.9 + M_\mathrm{I}  + 5\log h_{80} )\ ,\end{equation}
with $\mathrm{M_B}$ and $\mathrm{M_I}$ being the magnitudes corrected
to face-on for the B-band and I-band respectively, and $H_0 =
h_{80}80\Hubbleunits$.
The conclusions of G95 were the same as those of T98, with a fit
described by equation \ref{att_ratio}, and a luminosity dependence
given by
\begin{equation}
\label{gammaI_giovanelli}
\gamma _\mathrm{I}  =  - 0.16(15.25 + M_\mathrm{I}).
\end{equation}
However, M03 concluded that a simple linear law as defined in
equation \ref{att_ratio} was not adequate for describing the
dependence of attenuation on inclination observed in their sample. Thus,
M03 adopted a bilinear law fitting separately high and low
inclination systems as separated by $\log (a/b) = 0.5$. Looking for
even less linear shapes that better fit their observed colour
dependence on inclination, M03 also tried quadratic fits of the form
\begin{equation}
\label{quadratic}
A_{\lambda \upsilon } {\rm{ =  }}B_{\lambda \nu }  + C_{\lambda \nu } \log (a/b) + D_{\lambda \nu } \left[\log (a/b)\right]^2 \ ,
\end{equation}
where $C_{\lambda \nu }, B_{\lambda \nu}$, and $D_{\lambda \nu}$ are
constants that best fit their observed $\lambda-\nu$ colour dependence
on axial ratio.

\subsection{Comparison of Results with Observations}

First, the relative attenuation between the K-band and some other band
was examined. As mentioned in the last sub-section, T98 assumed that
the extinction effects on the K-band, although small, were not
entirely negligible. They assumed that, independent of inclination,
$\mathrm{A_K/A_B}$ relative to face-on was 0.15, and
$\mathrm{A_K/A_I}$ was 0.25. Our results for these ratios with both
the X99 and Z95 gradients are given in Table \ref {ratios}.

\begin{table}
\begin{minipage}{80mm}
\caption{Ratios of Relative to Face-on Attenuations.} 
\label{ratios}
\begin{tabular}{ccccc}
\hline 
&
\multicolumn{2}{c}{$A_\mathrm{K}/A_\mathrm{I}$} &
\multicolumn{2}{c}{$A_\mathrm{K}/A_\mathrm{B}$} \\

{$\theta $(deg)}& {X99}&{Z94}&{X99}&{Z94}\\
\hline
0&0&0&0&0 \\35.4&0.30&0.42&0.09&0.20 \\50.7&0.30&0.40&0.14&0.21\\63.2&0.39&0.35&0.21&0.20 \\74.4&0.34&0.35&0.20&0.23 \\85&0.38&0.39&0.26&0.28 \\90&0.41&0.43&0.30&0.32 \\ \\Average&0.39&0.33&0.20&0.24\\
\hline
\end{tabular}

\medskip
The ratios of relative to face-on attenuations between the K, I and B bands for the Xilouris and Zaritsky gradients at different inclinations for the Sbc galaxy model.
\end{minipage}
\end{table}

Table \ref{ratios} shows that the average relative attenuations found
in our simulations when the Xilouris gradient is assumed (0.20 and
0.39 for $A_\mathrm{K}/A_\mathrm{B}$ and $A_\mathrm{K}/A_\mathrm{I}$
respectively) are just a little higher than the ratios assumed by
T99. However, these ratios vary significantly and in a non-trivial
manner with inclination. This was not considered by T98. If this fact
is taken into account, their values of observed attenuations should be
about 15\% higher for large edge-on galaxies. For smaller galaxies and
intermediate inclinations the difference is not as large.

Figure \ref{incplots} shows a direct comparison of the attenuation
dependence on inclination and on luminosity with observations. For the
luminous galaxies, the Sbc variants, the Sc, and the G3, the Xilouris
gradient results in edge-on attenuations which are roughly consistent
with the observations. However, this gradient is too steep for the
low-luminosity galaxies, producing too little attenuation compared to
observations, and X99 only studied large spirals. For this reason, the
metallicity gradient for the G2, G1 and G0 galaxies was taken from the
results of MA04. In some cases, notably the G3, G2 and G1, the
no-gradient simulations seem to match the observations best, but as
galaxies are observed to have metallicity gradients, this model is not
acceptable and we chose the simple alternative of using the X99
gradient for the luminous galaxies and the MA04 gradient for the
smaller ones.

\begin{figure*} \begin {center} 
\includegraphics[width=0.81\textwidth]{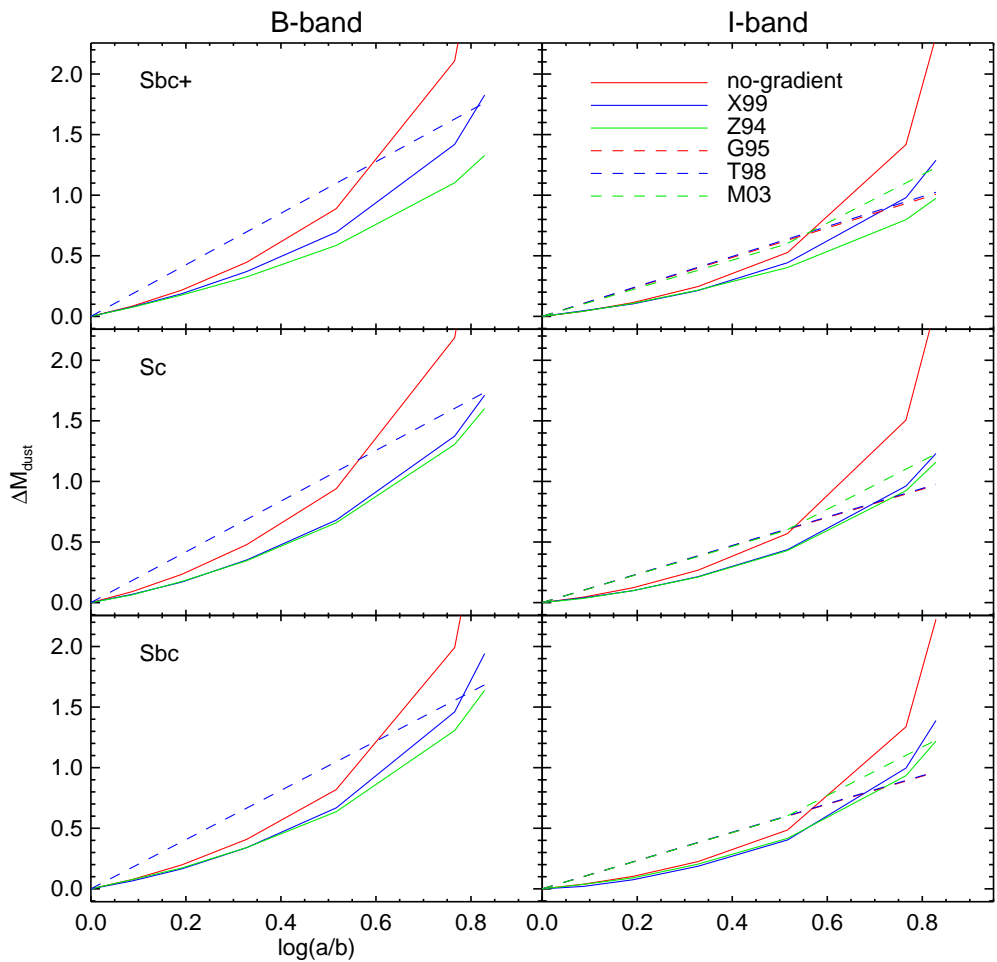}
 \end {center} 
 \caption{ Plots of attenuation relative to face-on vs. the $\log$ of
   the axial ratio, in the B-band (left column) and the I-band (right
   column), for our three most luminous galaxy models: the Sbc+, Sc,
   and Sbc (top to bottom). The solid lines represent the simulations
   for two metallicity gradients, Z94 (green) and X99 (blue), and a
   run with no gradient (red). Dashed lines are the best fits to the
   observations of G95, T98 and M03. The green dashed line is the
   bilinear line adopted by M03, valid only for their most luminous
   galaxies, while the rest are given by equations \ref{gammaB},
   \ref{gammaI} and \ref{gammaI_giovanelli} varying with
   luminosity. Although it is not evident if one of the gradients
   matches better the amplitude of the observations in I-band, it is
   noticeable that the X99 gradient matches the edge-on attenuations
   better than the Z94 gradient in the B-band (especially considering
   that T98 observations should be about 15$\%$ higher due to
   $\mathrm{K}$-band attenuation), hence we adopt it as our preferred
   gradient for the most luminous galaxy models. Our simulations
   evidently do not match the shape of the fits adopted by
   observers, as they predict less attenuation for intermediate
   inclinations than the observations show.}
\label{incplots} 
\end{figure*}

\begin{figure*} \begin
    {center}
\includegraphics[width=0.81\textwidth]{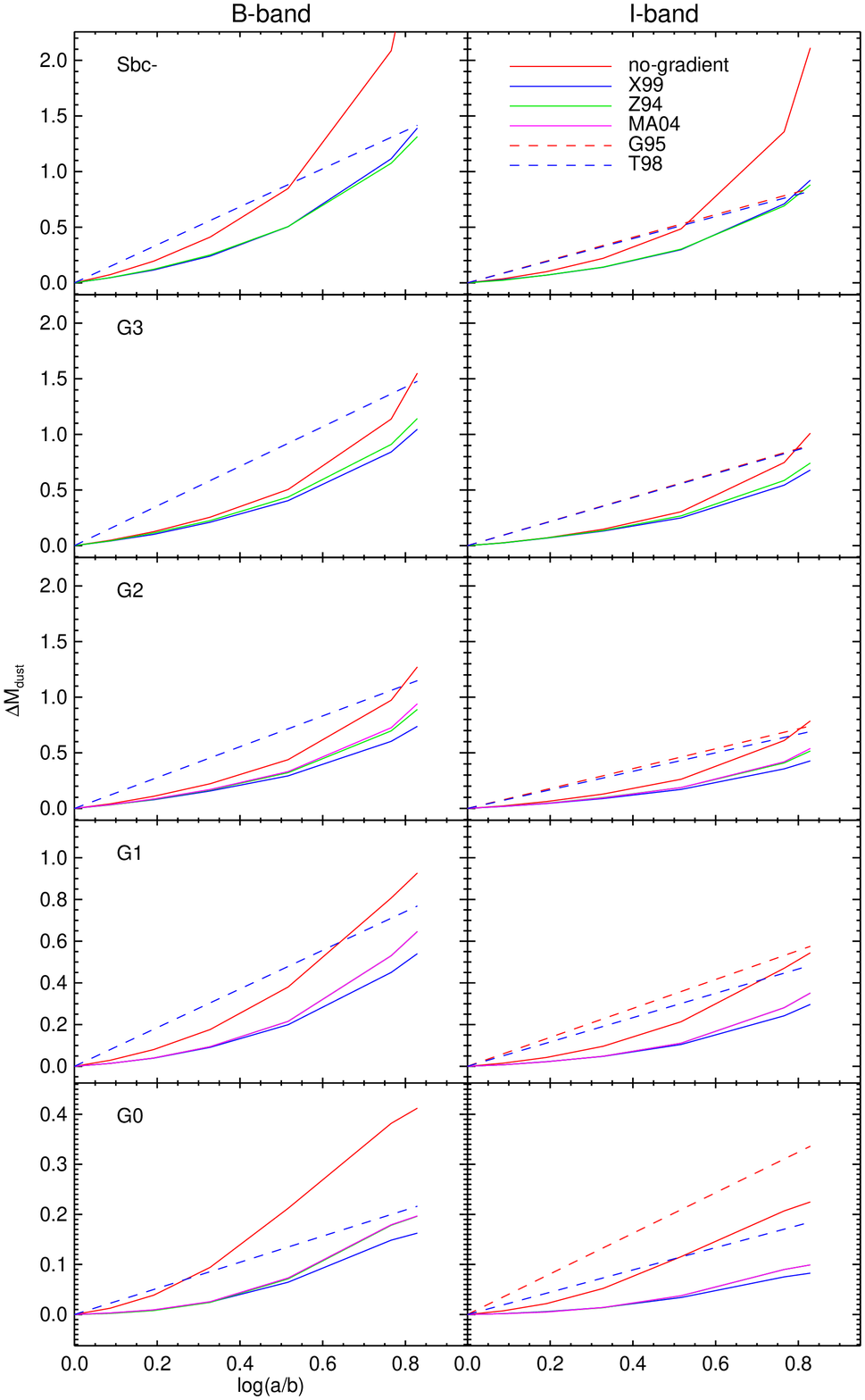}
  \end {center}  \contcaption{ The Sbc-, G3, G2,
    G1, and G0 (top to bottom). We can see how, for the G-series, the
    relative attenuation with the X99 (solid blue) gradient falls
    below the observations. Of the observed metallicity gradients,
    those of MA04 (solid pink) give a relative attenuation that is
    closest to what is observed, hence we have adopted these as our
    preferred gradients for the G0, G1, and G2. The results using the
    Z94 (solid green) for the G0, G1, and G2 are similar and overlap
    in some cases with those of MA04. For the lower-mass galaxies,
    using constant metallicity is sometimes the best-fitting
    model. Nevertheless, the no-gradient model was not adopted since
    no observations support such a metallicity distribution. }
\end{figure*} 

\begin{figure} \begin {center} \includegraphics[width=\columnwidth]{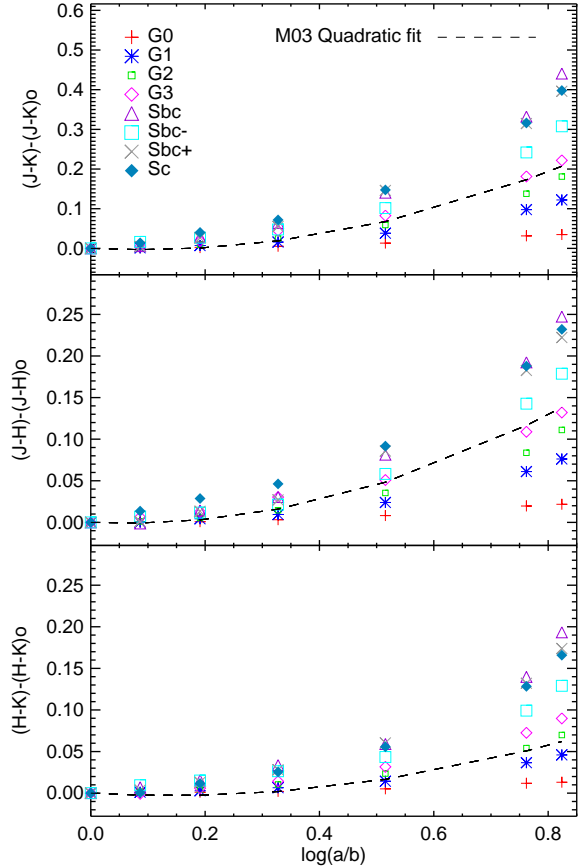} \end {center}  \caption{  The colours J-K, J-H, and H-K relative to face-on plotted against the log of the axial ratio. The black dashed lines represent the quadratic fits described by equation \ref {quadratic}. Over-plotted are the colour dependencies found in our models. We do not make a best fit to our points due to the fact that we have more bright galaxies than faint galaxies. The shape between simulations and observations is very similar, and a relation of the form given by equation \ref {quadratic} accurately describes the attenuation dependence on inclination in our simulations.} 
\label{quadratic_fig}
\end{figure}

It is apparent from Figure \ref{incplots} that, although the
simulations match some of the observed amplitudes, the dependence of
attenuation on inclination is certainly not linear; a relation of the
form $A_\lambda = \gamma_\lambda \log(a/b)$ is not appropriate. The
bilinear law adopted by M03 (dashed green in Figure \ref{incplots})
suggests a less linear relation, but still is not similar to what we
obtain in the simulations. We also compared the M03 quadratic fits
with our simulations. M03 (see Figure 4 in M03) used the statistical
average of the colour dependence on inclination observed in their full
sample to obtain the fits shown in Figure \ref{quadratic_fig}. In
order to compare with them, we observed the colour dependence on
inclination for each of our eight simulated galaxies, using the
adopted gradients for each (see Table \ref{testedgrads}); this is
shown in Figure \ref{quadratic_fig}. The shape of the quadratic fits
agrees rather well with the results of the simulations.

\begin{figure*} 
\begin {center}
  \includegraphics[width=0.9\textwidth]{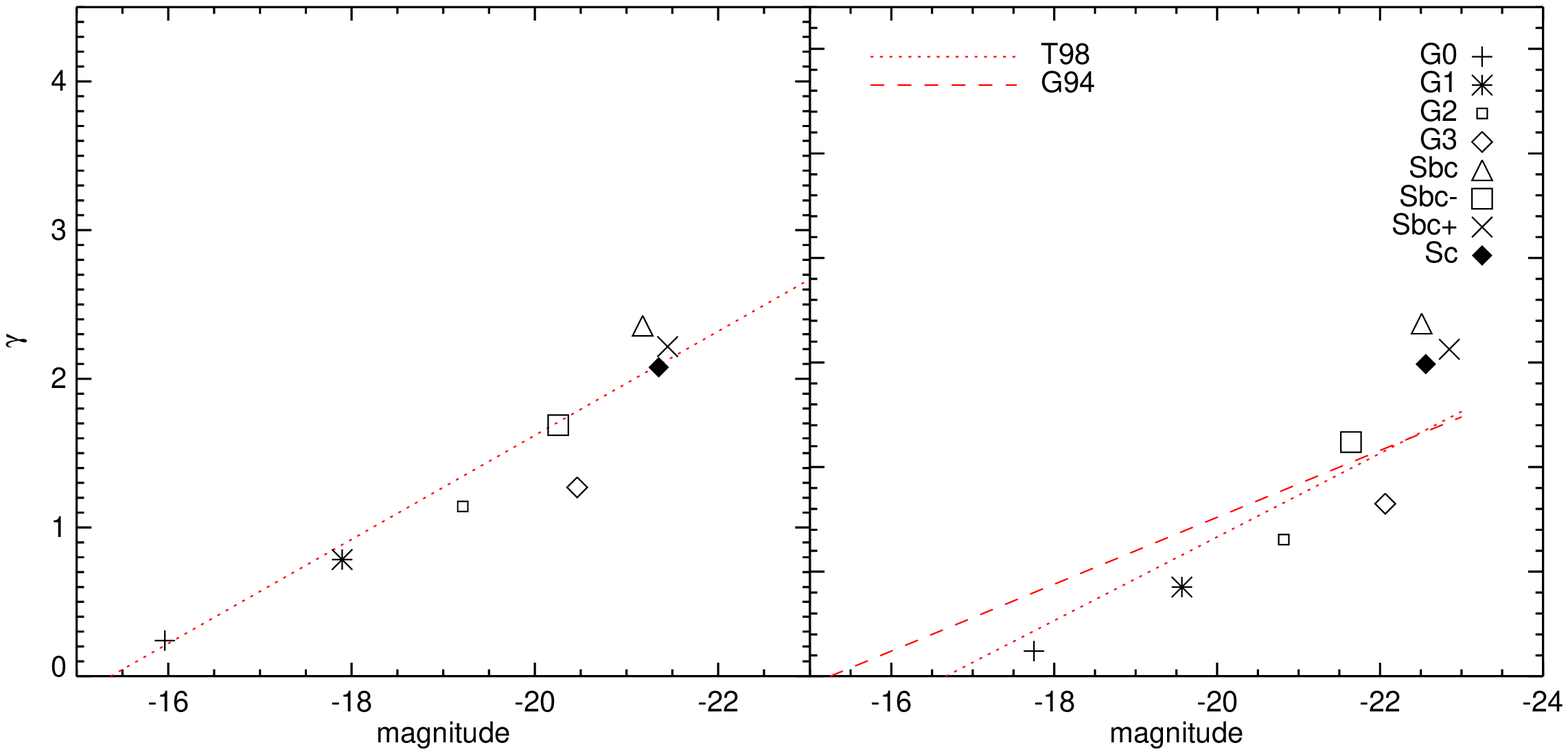} 
\end {center}
\caption{  The attenuation parameter $\gamma$, as defined by the
  relation \ref{att_ratio}, plotted against the face-on
  magnitudes of all our models in the B-band (left), and the I-band
  (right). The dotted and dashed lines are the best fits to the
  observations of T98 and G95 respectively. Our points match the
  I-band observations less well, but the agreement is still reasonable
  given the fact that we only have two galaxies
  fainter than $-20$.}
\label{lumdep}
\end{figure*}

Lastly, we compared the attenuation dependence on luminosity in our
galaxy models with what was observed by G95 and T98. T98 gives results
of luminosity dependence for both the I and B bands (equations
\ref{gammaB} and \ref{gammaI}), while G95 only gives results for the
I-band (equation \ref{gammaI_giovanelli}). This is shown in Figure
\ref{lumdep}, which illustrates that our points closely follow the
fits of G95 and T98, supporting the conclusion that the adopted
metallicity gradients are good for our galaxy models.  These gradients
and the face-on luminosities of our galaxy models when using these
gradients are given in Tables \ref{testedgrads} and \ref{lum}.

\begin {table}
\caption{Luminosities}
\label{lum} 
\begin{tabular}{ccccc}
\hline
Galaxy &  $\mathrm{M_B}$  & $\mathrm{M_I}$ & $\mathrm{M_K}$ & $L_{abs}/L_{bol}$ \\ 
\hline
Sbc+& -21.6 &-22.9 & -23.2 & 0.48 \\Sbc& -21.4  &-22.5 & -22.8 &0.44 \\Sc& -21.2 &-22.6& -22.6& 0.46 \\ Sbc-& -20.3&-21.6& -21.9&0.37 \\G3& -20.4&-22.0&-22.4&0.32 \\G2& -19.2 &-20.8&-21.2&0.32 \\G1&-17.9&-19.6&-19.9&0.14 \\G0&-16.0&-17.6&-18.1&0.02 \\ 
\hline
\end{tabular}

\medskip
Face-on luminosities in $\mathrm{B}$, $\mathrm{I}$, and $\mathrm{K}$ bands for
our set of galaxy models when the adopted metallicity gradients of
Table \ref{testedgrads} are used. The last column shows the fraction
of bolometric luminosity which is absorbed by dust.
\end{table}

So far we have been examining the attenuation in certain filter bands.
However, \mcrx \ calculates dust attenuation at all wavelengths, so we
can also calculate the total bolometric stellar luminosity that is
absorbed by dust and consequently re-emitted in the infrared.  When
mid/far-IR observations are available, this ``bolometric attenuation''
provides a sensitive measure of the total amount of dust, thus
avoiding the problem of determining the face-on attenuation.

\begin{figure}
\begin{center}
\includegraphics[width=\columnwidth]{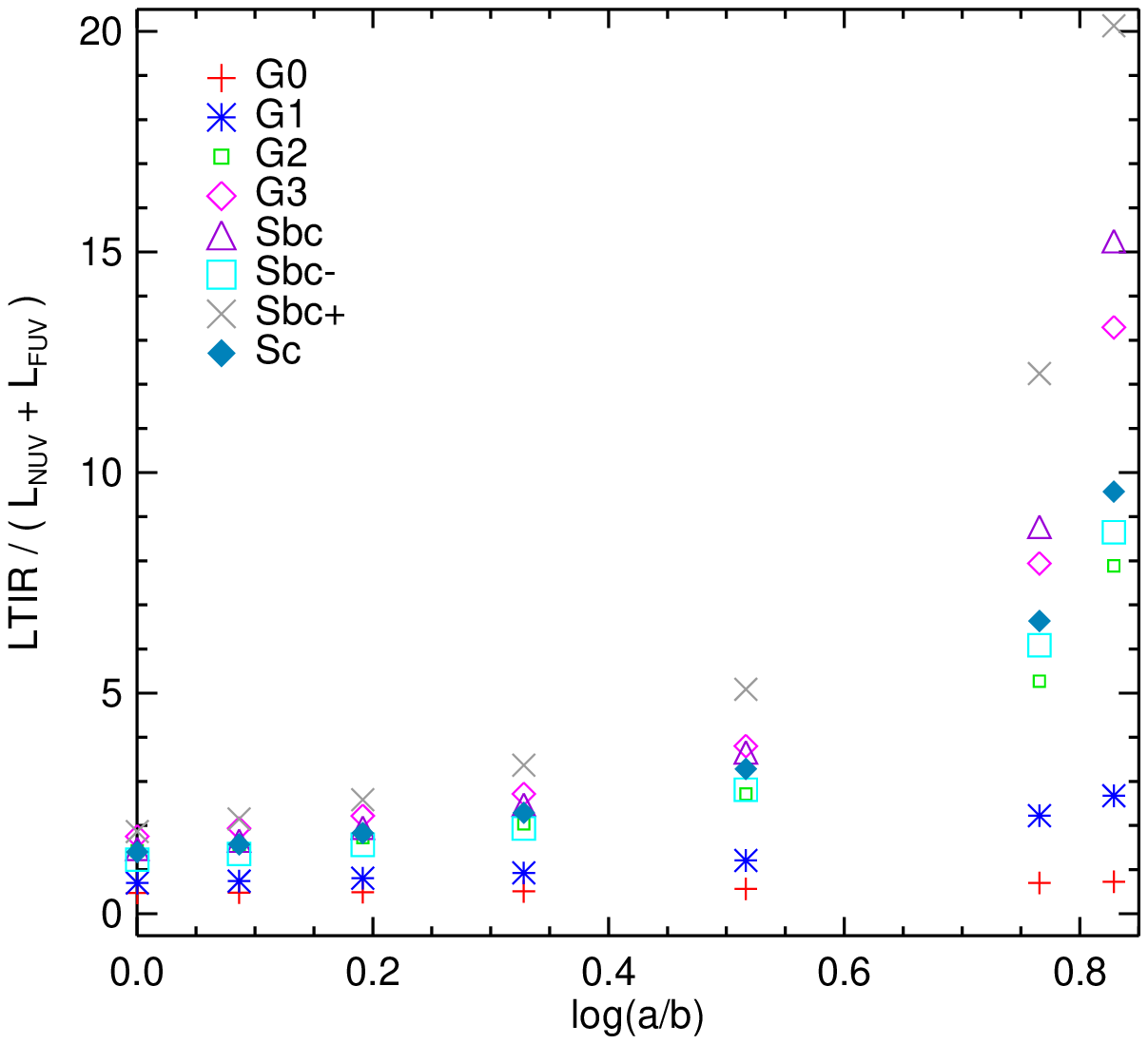}
\end {center}  
\caption{ Infrared-to-ultraviolet ratios for our models, as a function
  of inclination. This figure can be directly compared to Figures~4 \&
  5 in D07. Compared to the SINGS galaxies of the corresponding
  morphological types, the simulations show slightly lower dust
  attenuations but cover roughly the same range of values.  The total
  infrared emission in our models thus appears to be consistent with what
  is observed by SINGS.}
\label{irtouv}
\end{figure}

In order to compare with the Spitzer Infrared Nearby Galaxies Survey
(SINGS) sample \citep[][hereafter D07]{Daleetal07}, we have calculated
the infrared-to-ultraviolet ratios in our models in the same manner as
they did, by dividing the total infrared luminosity (TIR) with the
ultraviolet luminosity given by the sum of the luminosities at 1500
\AA \ and 2300 \AA \ (i.e. the FUV and NUV filter bands of the GALEX
satellite). Figure \ref{irtouv} shows the results for our different
models at different inclinations. We only compare within the range of
inclinations included in the D07 sample (i.e. 0 to ~70
degrees). Comparing the range of infrared-to-ultraviolet ratios for
the Sbc, G3 (which approximates an Sb type), and Sc models with the
galaxies of these types observed by D07 (see Figure 5 in D07), the
simulated galaxies show slightly lower attenuation. While D07 observed
ratios from about 1 to 10, we obtain ratios from about 1.5 to 7, but
the SINGS sample only contains a few star-forming galaxies with these
morphological types.  Overall, the amount of dust attenuation in our
galaxies, as given by the infrared-to-ultraviolet flux ratio, appears
to be fairly close to what is observed in real galaxies.

Another measurement of the bolometric attenuation was done by
\citet{Popescutuffs02} using the ISOPHOT instrument on the ISO
satellite, who found correlations between the overall attenuation and
the Hubble type of a sample of galaxies in the Virgo cluster. They
obtained a mean bolometric attenuation in galaxies of type Sab-Sc of
about 0.23.  Comparing the values in Table~\ref{lum} for the G3, Sbc-,
Sbc, Sbc+, and Sc to the results of \citet{Popescutuffs02}, the
simulations have attenuations that are about $15-20\%$ higher. (The
simulated galaxies here have slightly lower bolometric attenuations
than those previously obtained by \citet{pjetal05attn} using
simulations with constant metallicity.)

Apparently, the results of D07 and \citet{Popescutuffs02} are mutually
inconsistent. It is possible that the ISOPHOT observations, only
sensitive to wavelengths longer than $60\um$, missed a significant
fraction of the dust luminosity at shorter wavelengths.  Another
possibility is that the spiral galaxies in the Virgo cluster have
systematically lower gas content and lower star-formation rates than
the average field spiral.  Overall, the fact that our galaxies have slightly
lower bolometric attenuation compared to the newer and higher-quality
data of the SINGS galaxies, as well as slightly lower values of
$A_\lambda$, especially at intermediate inclinations, paints a
consistent picture that the simulations have dust attenuations on the
slightly low side of observed galaxies of the same morphological types.

\section{Comparison with Other Models}

\begin{figure}
\begin{center}
\includegraphics[width=\columnwidth]{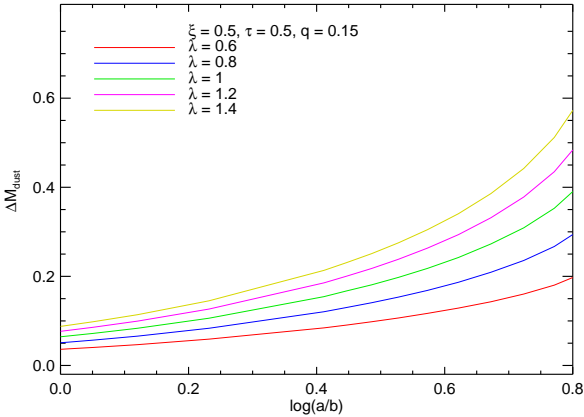}
\end {center}  
\caption{Attenuation relative to face-on vs. the $\log$ of the axial
  ratio for the ``quadruple exponential'' model when different values
  of $\lambda=R_d/R_s$ are given. We can see that the different values
  of $\lambda$ result in different magnitudes of edge-on attenuations,
  but do not affect the quadratic shape of the relation much.  }
\label{qua_lamb}
\end{figure}

A simple way to model dust attenuation in the disk of galaxies is to
perform radiative transfer calculations in smooth exponential
distributions of stars and dust, disregarding scattering. This is
unrealistic but helps to understand the behaviour of attenuation when
changing parameters such as inclination or dust-to-stellar
scalelengths/scaleheights. \citet{Disneyetal89} developed a triple
exponential model in which a single scalelength describes the radial
distribution of stars and dust, and two vertical scaleheights, one for
dust and one for stars, describe the vertical distributions of the
model. This same model was later used and extended by
\citet{Giovanellietal94} and M03. Motivated by the results of X99, we
explored the same model but allowing different scalelengths for dust and
stars (a ``quadruple exponential'' model). In a model like this, the
emissivity and absorption coefficients of the radiative-transfer
equation would be given by
\begin {eqnarray}
\label {quaplexeq}
\epsilon (r,z) & = & \epsilon _o \exp \left( - \frac{r}{{R_s }} -
  \frac{\left| z \right|}{Z_s }\right)\ \mbox{and} \\
 \kappa (r,z) & = & \kappa _o \exp \left( - \frac{r}{{R_d }} - \frac{{\left| z
  \right|}}{{Z_d }}\right) \ ,
\end{eqnarray}
with $\epsilon_o$ and $\kappa_o$ being the emissivity and the absorption
coefficient at the centre, respectively, $R_s$ and $Z_s$ the
scalelength and scaleheight of the stars, respectively, and $R_d$ and
$Z_d$ the scalelength and scaleheight of the dust, respectively. Figure
\ref{qua_lamb} plots the resulting attenuation dependence on
inclination for different values of $\lambda=R_d/R_s$, and with $\xi
=Z_d/Z_s=0.5$, $q=Z_s/R_s=0.15$, and $\tau=1$, where $\tau$ is the
central face-on optical depth
\begin {equation}
\label {quaplexeq2}
\tau  = \int {\kappa (0,z)} \,\mathrm {d}z = 
2\kappa _o \int_0^\infty  {\exp ( - \frac{z}{{Z_d }}} )\,\mathrm {d}z = 
2\kappa _o Z_d.
\end{equation}
\citet{Disneyetal89}, \citet{Giovanellietal94}, and M03 show how this
simple model behaves when different values of $\xi$, $q$, and $\tau$
are given. We find that when different values of $\lambda$ are assumed
the edge-on attenuation varies significantly, but the shape of the
curve remains the same (see Figure \ref{qua_lamb}).

More sophisticated radiative transfer models where anisotropic
scattering is considered and different parameterisations and spatial
distributions of clumps are included can be found extensively in the
literature (see the introductions of \citet{Pierinietal04} and
\citet{Tuffsetal04} for nice reviews of the history of these models),
but few explore the consequences of structure such as spiral arms in
the resulting attenuations \citep{corrandietal96, Semionovetal07}. If we
compare our results for the attenuation dependence on inclination
(Figure \ref{incplots}) with some of these models,
i.e. \citet{Giovanellietal94, bfg96, Bianchietal00, Ferraraetal99,
BaesDejonghe01, Byunetal94, Tuffsetal04} we cannot see any significant
difference that can be attributable to the geometry of our simulated
galaxies.

\begin{figure*} \begin {center}
\includegraphics[width=0.9\textwidth]{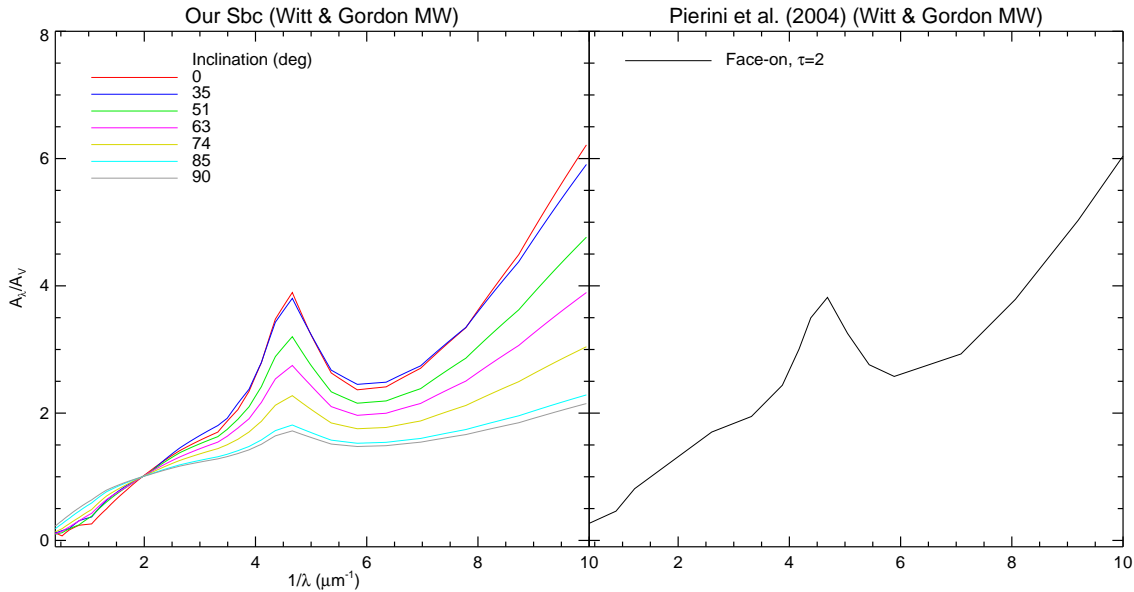}
 \end {center}  \caption{ Normalised attenuation curves for our Sbc model when the dust model adopted by \citet{Pierinietal04} is used in our simulation (left), and for the P04 homogeneous face-on model with $\tau = 2$ \ (right). We can see that there are no significant differences that can be related to the structure resolved in our hydrodynamic simulations. Figure 18 of P04 shows the dependence of their model with inclination, but only for $\tau = 1$, thus we have not plotted these results here. } 
\label{multi_att}
\end{figure*}

The dependence of attenuation on wavelength in our Sbc model (which
has a central optical depth $\tau$ close to 2) has been compared with
that of \citet[][hereafter P04]{Pierinietal04}, when we adopt their
model (see Table 1 of P04). The results are plotted in Figure
\ref{multi_att}. We did not find any significant differences between
the attenuation curves in P04 and ours when using the same dust model,
suggesting that the presence of spiral arms or any structure of
similar size in our simulations does not affect the attenuation
significantly.

It is important to notice that none of the above models, for values of
dust-to-star scale length/height ratios and central optical depths
normally observed in spiral galaxies, show attenuations that vary
linearly as a function of $\log (a/b)$. \citet{Ferraraetal99} found
that only values of dust-to-star scaleheight ratios of about 2.5
resulted in an attenuation dependence on inclination comparable to
that observed by \citet{devaucouleurs91}, but scaleheight ratios of
this magnitude are far from what observations suggest.  Thus, the
disagreement regarding the behaviour of the attenuation as a function
of inclination is not limited to our models and it is clear that
further work is needed to understand this discrepancy.

In the near future, the images that have been generated with \mcrx \
for this work will be analysed in order to quantify the changes in the
apparent scalelengths, central surface brightness, and axis ratios due
to the presence of dust, in a similar manner to what was done by
\citet{Mollenhoffetal06}.

\section{Summary and Conclusions}

\subsection{Summary}

Simulations of dust attenuation were performed in eight different SPH
simulations of isolated spiral galaxies, assuming different
metallicity gradients. The three metallicity gradients explored were
$-0.05$ dex/kpc, as suggested by Z94, the metallicity gradient
required to obtain a dust scalelength that is 1.4 times greater than
the stellar scalelength, as was observed by X99, and metallicities
obtained from the dependence of the metallicity gradient on luminosity
observed by MA04. We compared the dependence of attenuation on
inclination resulting from these gradients with observations by G95,
T98, and M03, finding that our simulations do not follow a linear
dependence on $\log(a/b)$. The non-linear fits adopted by M03 from
their observations of colour dependence on inclination were also
compared to the results of the simulations, giving a more satisfactory
match to the shape. The preferred gradients adopted from the above
comparisons were used to obtain a relation between the relative
face-on to edge-on attenuation and luminosity, finding that it
followed a rather linear behaviour that closely matches the
observations of G95, and T98. Finally, we compare our results to
previous models of galactic disks where structure such as spiral arms
are disregarded, showing that the curves obtained from our simulations
can be explained in terms of the same parameters (no parameters to
account for large structure resolved by our simulations are
necessary), and that the shape at intermediate inclinations is very
similar for both the models and our simulations.

\subsection{Conclusions}

1. When the radial metallicity gradient implied by the results of X99
is assumed, the observed amplitudes of relative attenuation from
face-on to edge-on in our most luminous galaxies agree fairly well
with the results of the simulations. The fainter galaxies of the set
required less steep gradients, and we adopted the gradients from MA04
for the G0, G1, and G2. This is not surprising considering that the
X99 results were obtained only for big spiral galaxies, and may not be
valid for smaller spiral galaxies.

2. Results for the ratios of attenuations between different bands in
Table \ref{ratios} show that there is always some attenuation in the
near infrared (K-band), with an inclination dependence such that the
attenuation in the K-band certainly is significant at high
inclinations. Our results show that the relative to face-on
attenuation in the K-band can be as high as 30{\%} of the attenuation
in the B-band when the Sbc model is seen edge-on, enough to introduce
bias of about 15{\%} if this is neglected.

3. Reproducing observed edge-on attenuations and their dependence on
luminosity does not imply that the simulations correctly describe the
extinction effects across all inclinations. The disagreement in the
shape of the attenuation vs. inclination relations between
observations and simulations suggests that either a simple linear
extinction law of the form $A_\lambda = \gamma _{\lambda }\log(a/b)$
is only a rough approximation to real galaxies, or that our
simulations are not good models of observed galaxies. It is difficult
to know whether the linear relations given by G95 and T98 are the
result of uncertainty in the observations, or if the relations
obtained in the simulations really are excluded by the data. The most
recent study by M03, which also has the largest sample, suggests that
a linear extinction law is not appropriate, and they propose both a
bilinear law and a quadratic law. The quadratic fits obtained by M03
have the same shape at intermediate inclinations as our
simulations. We have too few simulated galaxies to perform an adequate
statistical fit as has been done by M03, hence we cannot adequately
compare with the amplitude given by these quadratic fits. However, we
note that a good way to describe the attenuation dependence on
inclination found in the set of simulations performed here is by a
quadratic law of the form
\begin{equation}
\label{eq1}
\Delta \mathrm{M} = B+C\log (a/b)+D\left[\log (a/b)\right]^2,
\end{equation}
as proposed by \citet{Mastersetal03}. Other models obtain a shape
similar to ours, so this disagreement is not unique to our simulations

4. The bolometric attenuations in our simulations as measured by the
infrared-to-ultraviolet flux ratio are consistent with, but possibly
on the low side of, the range observed for galaxies of corresponding
morphological type in the SINGS sample by D07. However, the bolometric
attenuations in the simulations are significantly larger than those
obtained in an older study by \citet{Popescutuffs02}. This indicates
that the two observational studies are in conflict, regardless of
the uncertainty of the simulated results, as it will be difficult to
match both results simultaneously.

5. The attenuation vs. inclination relation obtained from models that
do not take account of spiral structure is very similar in shape (a
quadratic relation) to our calculations using hydrodynamic galaxy
models, and there is no evidence to conclude that the inclusion of
complex geometries like spiral arms makes a difference in the
attenuation as a function of inclination nor wavelength. Of course,
this is not true for highly disturbed systems like merging galaxies.

\medskip

This work was supported by programs HST-AR-10678 and HST-AR-10958,
provided by NASA through grants from the Space Telescope Science
Institute, which is operated by the Association of Universities for
Research in Astronomy, Incorporated, under NASA contract NAS5-26555.

This research used computational resources of the National Energy
Research Scientific Computing Center (NERSC), which is supported by
the Office of Science of the U.S.  Department of Energy, and of the
NASA Advanced Supercomputing Division (NAS).

\bibliographystyle{mn2e}
\bibliography{miguels}

\end{document}

%% file: natlatex_macros.tex
\newcommand{\kps}{\, {\rm km}/{\rm s}}

\newcommand{\um}{\, {\mu \rm m}}

\newcommand{\Hubbleunits}{\, {\rm km}/{\rm s}/{\rm Mpc}}
\newcommand{\kpc}{\, {\rm kpc}}

\newcommand{\Msun}{\,\mathrm{M}_{\odot}}

\newcommand{\mcrx}{\textit{Sunrise}}
\providecommand{\ga}{\gtrsim}
\providecommand{\la}{\lesssim}
\providecommand{\hii}{\mbox{H\,{\sc ii}}}